\documentclass[aps,pra,superscriptaddress,reprint]{revtex4-2}
\usepackage{amsmath}
\usepackage{amssymb}
\usepackage{mathrsfs}
\usepackage{graphicx}
\usepackage{booktabs}  %画表格用
\usepackage{multirow}  %画表格用
\usepackage[colorlinks=true, letterpaper=true, pdfstartview=FitV, linkcolor=blue, citecolor=blue, urlcolor=blue]{hyperref}

\begin{document}
	
\title{Weak anharmonicity sensing by single- and two-photon drives}
	
\author{Shi-Yu Zeng}
%\email{shiyuzeng@hust.edu.cn}
\affiliation{Key Laboratory of Low-Dimensional Quantum Structures and
Quantum Control of Ministry of Education, Key Laboratory for Matter
Microstructure and Function of Hunan Province, Department of Physics and
Synergetic Innovation Center for Quantum Effects and Applications, Hunan
Normal University, Changsha 410081, China} 
\author{Chao-Qun Ai}
%\email{1963485972@qq.com}
\affiliation{Key Laboratory of Low-Dimensional Quantum Structures and
Quantum Control of Ministry of Education, Key Laboratory for Matter
Microstructure and Function of Hunan Province, Department of Physics and
Synergetic Innovation Center for Quantum Effects and Applications, Hunan
Normal University, Changsha 410081, China} 
\author{Xi Cheng}
%\email{3261551697@qq.com}
\affiliation{Key Laboratory of Low-Dimensional Quantum Structures and
Quantum Control of Ministry of Education, Key Laboratory for Matter
Microstructure and Function of Hunan Province, Department of Physics and
Synergetic Innovation Center for Quantum Effects and Applications, Hunan
Normal University, Changsha 410081, China} 
\author{Ming-Wei Yan}
%\email{2021735576@qq.com}
\affiliation{Key Laboratory of Low-Dimensional Quantum Structures and
Quantum Control of Ministry of Education, Key Laboratory for Matter
Microstructure and Function of Hunan Province, Department of Physics and
Synergetic Innovation Center for Quantum Effects and Applications, Hunan
Normal University, Changsha 410081, China} 
\author{Xun-Wei Xu}
\email{xwxu@hunnu.edu.cn}
\affiliation{Key Laboratory of Low-Dimensional Quantum Structures and
Quantum Control of Ministry of Education, Key Laboratory for Matter
Microstructure and Function of Hunan Province, Department of Physics and
Synergetic Innovation Center for Quantum Effects and Applications, Hunan
Normal University, Changsha 410081, China} 
\affiliation{Hunan Research Center of the Basic Discipline for Quantum Effects and Quantum Technologies, Hunan Normal University, Changsha 410081, China}
\affiliation{Institute of Interdisciplinary Studies, Hunan Normal University, Changsha, 410081, China}

	\date{\today}
	
	\begin{abstract}
The optical cavity undergoes a quantum phase transition when the strength of a two-photon drive exceeds a critical point (CP), and the great sensitivity of CP in sensing has been recognized.
However, these methodologies are customized to sense linear perturbations, and quantum noise is divergent at the CP.
Here, we propose a scheme for sensing the weak Kerr nonlinearity in an optical cavity by both single- and two-photon drives, based on the CP for phase transition. 
We show that the mean photon number around the CP induced by the two-photon drive sensitively depends on the Kerr coefficient in the optical cavity, so that the weak anharmonicity in the optical cavity can be measured sensitively by detecting the mean photon number.
Moreover, we demonstrate that the single-photon drive provides an effective way to suppress the quantum noise and improve the signal-to-noise ratio. 
This scheme can be applied to detecting the weak nonlinear interactions in a wide range of optical systems.
	\end{abstract}
	
	\maketitle
	
	\section{Introduction}
	
Sensors are the fundamental devices in the modern world, with wide applications in physics~\cite{Gil-Santos2010,DeMille2024NatPh,Rovny2024NatRP}, chemistry~\cite{Cimini2019PRA,Sarkar2024SciA}, biology~\cite{He2011,Aslam2023NatRP} and other areas.
Many new mechanisms are proposed to enhance the sensitivity of sensors, and a noticeable one is the sensing schemes based on the singularity points, such as exceptional points (EPs) in non-Hermitian systems and critical point (CP) for quantum phase transition.
EPs are the unique degeneracy points in the non-Hermitian systems~\cite{Bender1998,GuoA2009PRL,Ruter2010NatPh,El-Ganainy2018,PengB2014NatPh,ChangL2014NaPho,SongP2024NatCo} that the eigen-values and eigen-modes coalesce simultaneously and the system is highly sensitive to small perturbations near the EPs~\cite{Ozdemir2019NatMa,miri2019exceptional}.
For a \(n\)-th order EP, when the system undergoes a small perturbation with a strength of $\varepsilon$, the system's energy will exhibit a strong divergence as \(\epsilon^{1/n}\), which have been demonstrated theoretically and experimentally~\cite{Wiersig2014,Hodaei2017Natur,ChenW2017Natur,Chenxiao2019,Hokmabadi2019Natur,LaiYH2019Natur,Kononchuk2022Natur,Zhou2023NatCo,Ruan2025NaPho,Bai2023NSRev}.
In contrast, the CP enhanced sensing is induced by the divergent susceptibility of physical systems near a CP for quantum phase transition, and such effect has also been demonstrated theoretically and experimentally~\cite{Beaulieu2025PRXq,Sarkar2025NatCo,DingDS2022NatPh,DiCandia2023npjQI,LvJH2022PRA,TangSB2023PRA,Garbe2020PRL,Raghunandan2018PRL}.
Nonetheless, most of these sensing schemes are customized to sense linear perturbations, and one would like to explore new sensing techniques for the detection of anharmonic perturbations.

Anharmonicity is an ubiquitous phenomenon in a wide range of optical systems induced by light-matter interactions, such as the dispersive atom-resonator interaction~\cite{Haroche2006}, magnon-photon interaction~\cite{ZARERAMESHTI20221}, and the optomechanical interaction~\cite{Aspelmeyer2014RMP}. But often it is weak and difficult to be accurately measured.
Recently, a scheme for enhanced sensing of weak anharmonicities was proposed based on the coherence between two modes induced by a common vacuum~\cite{Nair2021}.
It was shown that there is a singularity point in the two modes with the common reservoir that distinguishes an especially long-lived resonance, and it leads to a tremendous buildup in the steady-state amplitudes and the nonlinear response being highly sensitive to variations in the strength of anharmonicity.
After that, another scheme was proposed to enhance the sensitivity of a quantum system to nonlinearities by homodyning the amplitude quadrature of the cavity field~\cite{cui2023enhancing}.
Besides the dissipative coupling between the two cavity modes, one of cavity modes is subject to both single- and two-photon drives, so that the sensing protocol exhibits a even higher sensitivity.

Two-photon drive, that can be realized in a wide range of settings, is a subject extremely studied in quantum optics~\cite{Walls1994}.
Two-photon drive obtained by means of the degenerate parametric down-conversion is the key ingredient in achieving squeezed states of the electromagnetic field~\cite{WuLA1986PRL}. 
A single bosonic mode subject to two-photon driving can exhibit an EP without having dissipation and the corresponding driving noise~\cite{WangYX2019PRA}.
A Kerr nonlinear resonator with a two-photon drive provides a promising candidates for engineering quantum states and realizing elementary gates for universal quantum computation~\cite{Goto2016PRA,Puri2017npjQI,Grimm2020Natur,Iyama2024NatCo}.
The critical phenomena in a nonlinear oscillator subject to two-photon driving was analyzed in detail, and the rich physics were discovered, including a continuous phase transition, $Z_2$ symmetry breaking, $\mathcal{P}\mathcal{T}$ symmetry, state squeezing, and critical fluctuations~\cite{ZhangXH2021PRA}.
Besides, two-photon drive has also been applied to induce nonlinear interaction enhancement~\cite{LvXY2015PRL,QinW2018PRL,WangY2023SCPMA}, nonreciprocity~\cite{TangL2022PRL,ShenCP2023PRA,LuTX2024SCPMA}, and quantum sensing enhancement~\cite{Roy2021Opt,QinW2022PRL,ZhaoW2020SCPMA,Zhang2024OQ}.

%Despite the sensing schemes based on CP exhibiting high detection sensitivity, how to suppress the quantum noise around the CP is still an imperious issue for research.

In this paper, we propose a scheme for sensing the weak Kerr nonlinearity in an optical cavity by both single- and two-photon drives, based on the CP for phase transition.
Remarkably, we only need one optical mode, which is much simpler than the previous schemes for weak anharmonicity sensing~\cite{Nair2021,cui2023enhancing}, which are achieved based on two optical modes coupled directly by exchanging photons and indirectly by a common reservoir.
What's more, we show that the quantum noise at the CP can be suppressed by adjusting the strength of the single-photon drive, which provide an effective way to improve the signal-to-noise ratio (SNR) for sensing.

The remainder of our paper is organized as follows.
In Sec.~\ref{Model}, we introduce the model of the system and the CP for sensing.
In Sec.~\ref{NonlinearSensing}, we show the relation between the mean photon number and the strength of nonlinearity around the CP, and demonstrate the sensitivity of the mean photon number to the strength of nonlinearity.
In Sec.~\ref{NoiseAnalysis}, we discuss the effect of the quantum noise and how to suppress the quantum noise.
Finally, we give the conclusions of this work in Sec.~\ref{Concl}.
	
	\section{System Model}\label{Model}
	
\begin{figure}[tbp]
		\centering
		\includegraphics[bb=14 1 1603 540, width=8 cm, clip]{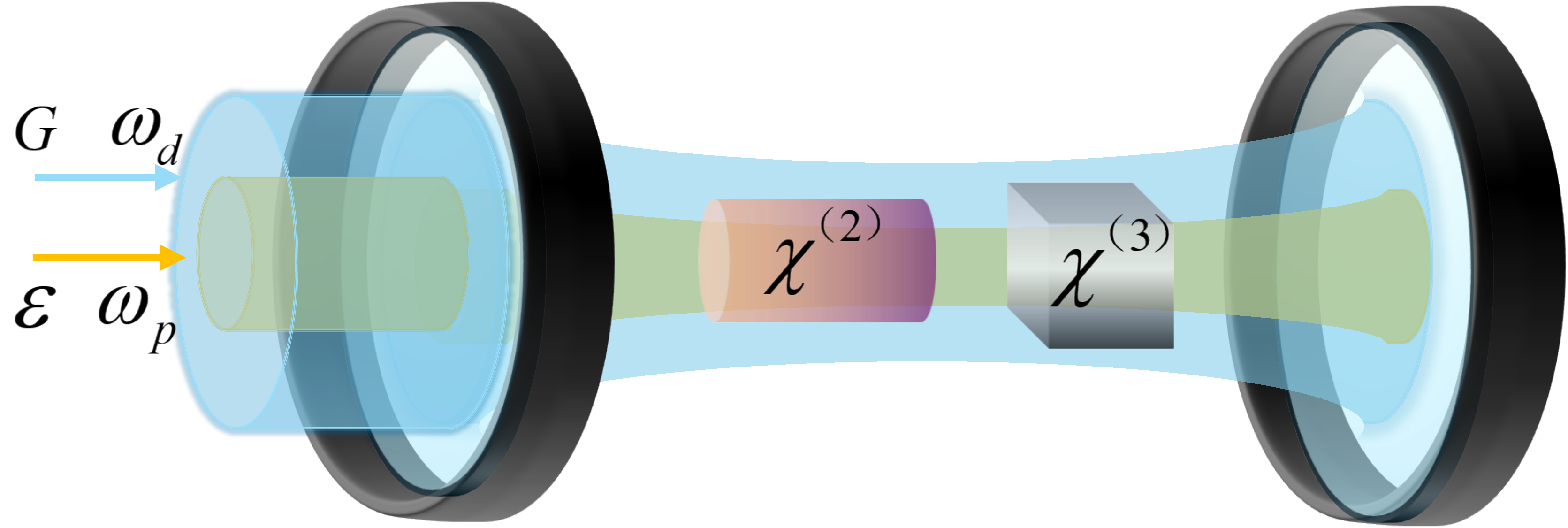}
		\caption{Schematic diagram of the system. An optical cavity is driven by both single-photon and two-photon drives. \( \varepsilon \) and \( \omega_p \) are the amplitude and frequency of the single-photon drive, respectively. A two-photon drive is achieved through a second-order nonlinear medium \( \chi^{(2)} \) in the cavity pumped with the amplitude \( G \) and frequency \( \omega_d \). The amplitude $U$ of the Kerr nonlinearity \( \chi^{(3)} \) in the cavity is the parameter for detection.}   
		\label{fig1}
	\end{figure}

	We consider an optical cavity mode subjected to both single-photon and two-photon drives, to measure the strength of weak anharmonicity in the optical cavity. The schematic diagram is sketched in Fig~\ref{fig1}. The Hamiltonian of the system can be expressed as (setting \(\hbar = 1\)):
	\begin{equation}
		\begin{aligned}
			 H_0= & \omega_a a^{\dagger} a+U a^{\dagger} a^{\dagger} a a+i \varepsilon\left(a^{\dagger} e^{-i \omega_p}-a e^{i \omega_p}\right) \\
			& +G\left(a^{\dagger 2} e^{-i \omega_d}+a^2 e^{i \omega_d}\right),
		\end{aligned}
	\end{equation}
	where $\omega_a$ is the frequency of the cavity mode with the annihilation and creation operators $a$ and $a^\dagger$, $U$ represents the Kerr coefficient, $\varepsilon$ is the amplitude of single-photon drive with frequency $\omega_p$. A two-photon drive is applied to the cavity with pumping amplitude $G$ and frequency $\omega_d$.	
Here, we assume that the pumping frequencies satisfy the conditions $\omega_p = \omega_d/2$.
In a frame rotating with the frequency $\omega_p$ of the single-photon drive, the Hamiltonian of the system is given by
	\begin{equation}\label{EqH}
H=\Delta a^{\dagger} a+U a^{\dagger} a^{\dagger} a a+i \varepsilon\left( a^{\dagger}-a \right)+G\left(a^{\dagger 2}+a^2\right),
	\end{equation}
where $\Delta=\omega_a-\omega_p$ is the detuning between the cavity mode and the single-photon drive field.

	Based on the Hamiltonian~(\ref{EqH}), the dynamics of the operators for the optical cavity system can be described by the quantum Langevin equations (QLEs)
\begin{eqnarray}
\frac{d a}{d t}= & -i \Delta a-\frac{\gamma}{2} a-i 2 G a^{\dagger}-i 2 U a^{\dagger} a a+\varepsilon+\sqrt{\gamma} a_{\mathrm{in}}, \\
\frac{d a^{\dagger}}{d t}= & i \Delta a^{\dagger}-\frac{\gamma}{2} a^{\dagger}+i 2 G a+i 2 U a^{\dagger} a^{\dagger} a+\varepsilon+\sqrt{\gamma} a_{\mathrm{in}}^{\dagger},
\end{eqnarray}
where the parameter \(\gamma\) is the dissipation rate of the optical cavity, and \(a_{\text{in}}\) is the operator of the noise input to the optical cavity with zero mean values, i.e., \(\langle a_{\text{in}} \rangle = 0\). To analyze the steady-state and fluctuation properties of the system, we linearize the QLEs by introducing \(a = \alpha + \delta a\), where \(\alpha \equiv \langle a \rangle\) is the mean value of the annihilation operator, and \(\delta a\) is the fluctuation operator of the annihilation operator with \(\langle \delta a \rangle = 0\). Then the mean values \(\alpha\) and  \(\alpha^*\) satisfy the semiclassical mean-field equations as
\begin{eqnarray}
\frac{d \alpha}{d t}= & \left(-\frac{\gamma}{2}-i\Delta'\right) \alpha-i 2 G \alpha^*+\varepsilon, \label{MF_eq1}\\
\frac{d \alpha^*}{d t}= & \left(-\frac{\gamma}{2}+i\Delta'\right) \alpha^*+i 2 G \alpha+\varepsilon,\label{MF_eq2}
\end{eqnarray}
where $\Delta'\equiv \Delta+2 U|\alpha|^2$, and the QLEs for the fluctuation operator $\delta a$ and $\delta a^{\dagger}$ are governed by
\begin{eqnarray}\label{QLE_f_eq1}
\frac{d\delta a}{dt} &=&\left( -\frac{\gamma }{2}-i\Delta''\right) \delta a-i2G'
\delta a^{\dag }-i2U\delta a^{\dag }\delta a\delta a  \nonumber \\
&&-i4U\alpha \delta a^{\dag }\delta a-i2U\alpha ^{\ast }\delta a\delta a+\sqrt{\gamma }a_{\mathrm{in}},
\end{eqnarray}%
\begin{eqnarray}\label{QLE_f_eq2}
\frac{d\delta a^{\dag }}{dt} &=&\left( -\frac{\gamma }{2}+i\Delta''\right) \delta a^{\dag }+i2G'^{*} \delta a+i2U\delta a^{\dag }\delta a^{\dag }\delta a  \nonumber \\
&&+i4U\alpha ^{\ast }\delta a^{\dag }\delta a+i2U\alpha \delta a^{\dag
}\delta a^{\dag }+\sqrt{\gamma }a_{\mathrm{in}}^{\dag },
\end{eqnarray}
where $\Delta''\equiv \Delta+4 U|\alpha|^2$ and $G'\equiv G+U\alpha ^{2}$. In the limit that $U/\gamma \rightarrow 0$, the mean-field equations [(\ref{MF_eq1}) and (\ref{MF_eq2})] can give the accurate results~\cite{ZhangXH2021PRA}.

	\begin{figure*}[tbp]
		\centering
		\includegraphics[bb=53  515  516 639, width=18cm, clip]{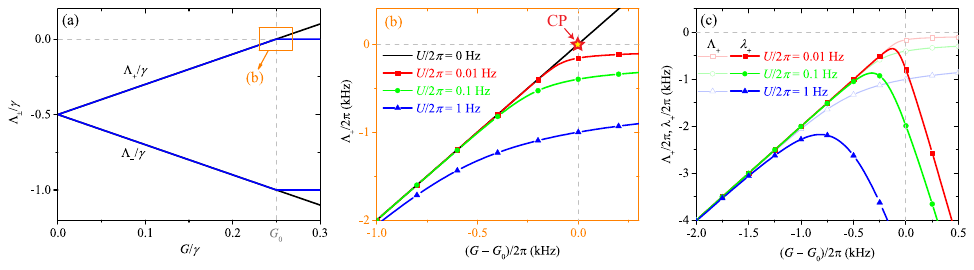}
		\caption{(a) The eigenvalues of the matrix \(\mathbf{M}\) (i.e., $\Lambda_{\pm}/\gamma$) versus the amplitude of two-photon drive $G/\gamma$. (b) A local enlargement of (a) in the region near the CP for different Kerr coefficient $U/2\pi$ (0 Hz, 0.01 Hz, 0.1 Hz, 1 Hz). (c) The eigenvalues of $\Lambda_{+}/2\pi$ and $\lambda_{+}/2\pi$ versus $(G-G_0)/2\pi$ near the CP for different Kerr coefficient $U/2\pi$. The parameters used are \( \gamma / 2\pi = 1 \times 10^9 \) Hz, $\Delta=0$, and \( \varepsilon / \gamma = 10^{-3} \).}   
		\label{fig2}
	\end{figure*}

The mean-field equations can be expressed in matrix form as
	\begin{equation}
		\frac{d}{d t}\binom{\alpha}{\alpha^*}=\mathbf{M}\binom{\alpha}{\alpha^*}+\binom{\varepsilon}{\varepsilon},
	\end{equation}	
where
	\begin{equation}
	\mathbf{M} =
	\begin{pmatrix}
		-\frac{\gamma}{2} - i\Delta' & -i2G \\[8pt]
		i2G & -\frac{\gamma}{2} + i\Delta'
	\end{pmatrix}.
	\end{equation}	
The eigenvalues of the system matrix $\mathbf{M}$ are
	\begin{equation}
	\Lambda_{\pm} = -\frac{\gamma}{2} \pm \sqrt{4G^2 - \Delta'^2 },
	\end{equation}	
where the real parts of $\Lambda_{\pm}$ are related to the effective damping rate of the system, and the imaginary parts corresponds to the eigenfrequency. When \(\Delta' = 0\), the eigenvalues read
\begin{equation}
\Lambda_{ \pm}=-\frac{\gamma}{2} \pm 2 G.
\end{equation}
When the two-photon driving strength approaches the critical strength \(G_0 = \gamma / 4\), one of the real parts of $\Lambda_{\pm}$ tends to zero (\(\Lambda_+ = 0\)). This is the critical point (CP) for phase transition~\cite{ZhangXH2021PRA}, and the system becomes unstable when $\Lambda_{\pm}>0$.

	To discuss the stability of the system, we also need to analyze the eigenvalues of the linearized QLEs for the fluctuation operators. The linearized QLEs are obtained by neglecting the nonlinear terms in Eqs.~(\ref{QLE_f_eq1}) and (\ref{QLE_f_eq2}), and they are given by
	\begin{equation}
		\frac{d}{d t}\binom{\delta a}{\delta a^{\dag }}=\mathbf{M'}\binom{\delta a}{\delta a^{\dag }}+\sqrt{\gamma } \binom{a_{\mathrm{in}}}{a_{\mathrm{in}}^{\dag }},
	\end{equation}	
where
	\begin{equation}
	\mathbf{M'} =
	\begin{pmatrix}
		-\frac{\gamma}{2} - i\Delta'' & -i2G' \\[8pt]
		i2G'^{*} & -\frac{\gamma}{2} + i\Delta''
	\end{pmatrix}.
	\end{equation}
The eigenvalues of $\mathbf{M'}$ are shown as
	\begin{equation}
\lambda_{\pm} = -\frac{\gamma}{2} \pm \sqrt{4|G'|^2 - \Delta''^2 }.
	\end{equation}
The system is stable only if the real parts of all the eigenvalues of the matrices \(\mathbf{M}\) and \(\mathbf{M'}\) are negative.
Notably, the eigenvalues of the matrices \(\mathbf{M}\) and \(\mathbf{M'}\) depend on the value of $\alpha$.
	
We get the expression of $\alpha$ from the solution of Eqs.~(\ref{MF_eq1}) and (\ref{MF_eq2}) as
	\begin{eqnarray}\label{phontonEq}
		\alpha &=& -\frac{-\frac{\gamma}{2}+i \left(\Delta'+2 G\right)}{\Delta'^2+\left(\frac{\gamma}{2}\right)^2-4 G^2} \varepsilon \nonumber\\
&=& -\frac{-\frac{\gamma}{2}+i \left(\Delta'+2 G\right)}{\Lambda_{+}\Lambda_{-}} \varepsilon,
	\end{eqnarray}
in the steady state (i.e., \( d\alpha/dt = 0 \)).
As $\Delta'\equiv \Delta+2 U|\alpha|^2$, the value of $\alpha$ is correlated nonlinearly with the amplitude of single-photon drive $\varepsilon$, especially for $|\alpha|^2\gg 1$, which provide us an effective way to measure the Kerr coefficient $U$ (see next section).
What's more, $\alpha$ is in inverse proportion to the eigenvalues of the matrix \(\mathbf{M}\) (i.e., $\Lambda_{\pm}$), and $\alpha$ becomes divergent at the CP for $\Lambda_{\pm}=0$.

Let us discuss the relation between the eigenvalues of the matrix \(\mathbf{M}\) (i.e., $\Lambda_{\pm}$) and the amplitude $G$ of two-photon drive in detail. The eigenvalues $\Lambda_{\pm}$ versus $G$ are shown in Fig.~\ref{fig2}(a). It can be seen that with the enhance of the two-photon driving strength, the eigenvalue $\Lambda_{+}$ ($\Lambda_{-}$) increases (decreases) linearly, before reaching the critical driving strength \(G_0 = \gamma / 4\). In this regime, the effect of Kerr nonlinearity on the eigenvalues can be neglected. As driving strength approaching the critical driving strength $G_0$, $\alpha$ increases drastically and the effect of the Kerr nonlinearity becomes more and more significant. The eigenvalue $\Lambda_{+}$ versus the amplitude $G-G_0$ of the two-photon drive for different values of $U$ is shown in Fig.~\ref{fig2}(b). 
If there is no Kerr nonlinearity ($U=0$), the system becomes unstable for $\Lambda_{+}>0$ and $\lambda_{+}>0$ if $G>G_0$.
However, the perfect linear optical system ($U=0$) does not exist and the nonlinearity effect become noticeable when the photon number is very large.
In the vicinity of the CP, the nonlinear effects will suppress the system's photon number as well as the eigenvalue $\Lambda_{+}$. Under the suppression of the nonlinear effects,  $\Lambda_{+}$ is always less than 0.

	\begin{figure*}[tbp]
		\centering
		\includegraphics[bb=7 504 595 675, width=18cm, clip]{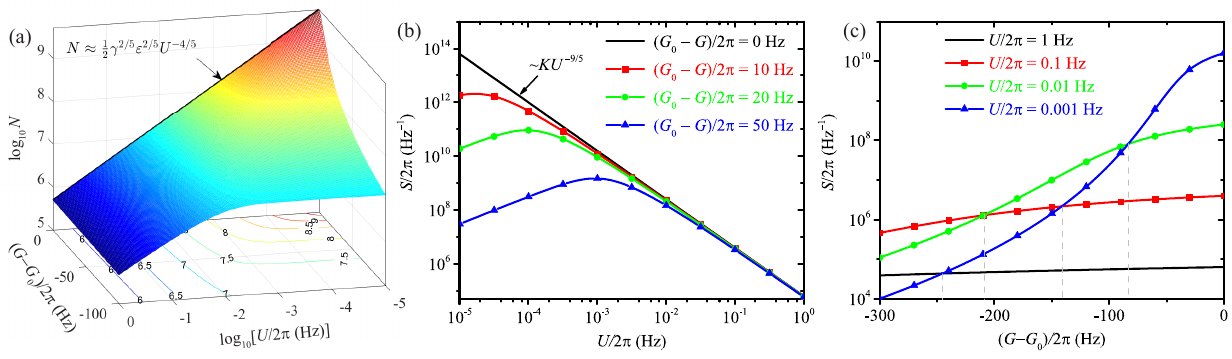}
		\caption{(a) The mean photon number $\log_{10}N$ versus the Kerr coefficient $\log_{10}[U/2\pi]$ and two-photon driving strength $(G-G_0)/2\pi$. The black line shows the photon number $N$ given in Eq.~(\ref{MeanPh}). (b) The sensitivity $S/2\pi$ versus the Kerr coefficient $U/2\pi$ for different two-photon driving strength $G$. (c) The sensitivity $S/2\pi$ versus the two-photon driving strength $(G-G_0)/2\pi$ for different Kerr coefficient $U$. The parameters used are the same as in Fig.~\ref{fig2}.}   
		\label{fig3}
	\end{figure*}

To make sure stability for the used parameters, we also discuss the relation between the eigenvalue of the matrix \(\mathbf{M'}\) (i.e., $\lambda_{+}$) and the amplitude $G$ of two-photon drive in Fig.~\ref{fig2}(c).
When $G\ll G_0$, we have $\lambda_{+}\approx \Lambda_{+}$. In the vicinity of the CP, we have $\lambda_{+} < \Lambda_{+}<0$, which means that the system is stable around the CP.
In the following discussions, we will make sure the conditions for stability are satisfied for the used parameters.

	\section{nonlinearity sensing}\label{NonlinearSensing}

As mentioned in the above section [i.e., Eq.~(\ref{phontonEq})] that the steady-state photon number $N=|\alpha|^2$ depends on the Kerr coefficient $U$, which inspires us to sense the weak anharmonicity in the optical systems by detecting the steady-state photon number. As shown in Fig.~\ref{fig2}(b), the difference of the eigenvalue $\Lambda_{+}$ for different Kerr coefficient $U$ becomes significant around the CP for $G=G_0$. Thus we will focus on the case around the CP. At the CP, the relation between the steady-state photon number and the Kerr coefficient $U$ is given by
	\begin{equation}\label{MeanPh}
		N \approx \frac{1}{2} \gamma^{2/5}\varepsilon^{2/5}U^{-4/5}, 
	\end{equation}
under the conditions \( \Delta = 0 \) and \( U|\alpha|^2 \ll G \).
It's worth mentioning that with the parameters selected in this paper, we have \( U|\alpha|^2 / G \ll 1\), which meets this approximate condition.
The sensitivity of the photon number $N$ to the Kerr coefficient $U$ can be defined by
	\begin{equation}
		S = \left| \frac{dN}{dU} \right|. 
	\end{equation}
At the CP ($G=G_0$), based on Eq.(\ref{sensitivity}), we can get the sensitivity as
\begin{equation}\label{sensitivity}
  S = K U^{-9/5},
\end{equation}
where
\begin{equation}
  K = \frac{2}{5}\gamma^{2/5} \varepsilon^{2/5}.
\end{equation}
We note that, for the scheme proposed in Ref.~\cite{Nair2021}, the sensitivity of the intensity to $U$ is encoded as $S\propto |U|^{-5/3}$, based on dissipatively coupled Anti-PT symmetric systems.
In addition, the sensitivity to nonlinearity is enhanced through homodyne detection in dissipatively coupled systems, and the sensitivity was shown as $S\propto |U|^{-2}$, in Ref.~\cite{cui2023enhancing}.

To show the relation between the photon number $N$ and the Kerr coefficient $U$ more intuitively, the mean photon number $\log_{10}N$ is plotted as a function of the Kerr coefficient $\log_{10}[U/2\pi]$ and two-photon driving strength $(G-G_0)/2\pi$ in Fig.~\ref{fig3}(a). At $G=G_0$, $\log_{10}N$ decreases linearly with the increase of $\log_{10}[U/2\pi]$, which agrees well with the analytical result given in Eq.~(\ref{MeanPh}) (black line).
As the increase of the difference $(G_0-G)$, $\log_{10}N$ drops quickly, which has a great impact on the sensitivity to nonlinearity.

The sensitivity to nonlinearity is plotted as a function of the Kerr coefficient $U/2\pi$ for different two-photon driving strength $G$ in Fig.~\ref{fig3}(b).
At the CP ($G=G_0$), the sensitivity $S$ decreases monotonically with the increase of $U$, by the scaling law $S\propto |U|^{-9/5}$, which agrees well with the analytical expression given in Eq.~(\ref{sensitivity}).
As $G<G_0$, the sensitivity $S$ increases first and then decreases with the increase of $U$. So there is a maximal sensitivity $S$ for $G<G_0$, and the maximal sensitivity $S$ decreases with the increase of the value of $G_0-G$.

To study the robustness of the high sensitivity against the deviates of the two-photon driving strength from the CP, we show the sensitivity to nonlinearity as a function of the two-photon driving strength $G-G_0$ for different Kerr coefficient $U/2\pi$ in Fig.~\ref{fig3}(c). The sensitivity $S$ decreases monotonically with the increase of deviates $G_0-G$.
It is worth to mention that the sensitivity $S$ is more sensitive (i.e., decrease faster) to the deviates $G_0-G$ for smaller Kerr coefficient $U/2\pi$.

	\section{Noise Analysis}\label{NoiseAnalysis}

	\begin{figure}[tbp]
		\centering
		\includegraphics[bb=137 195 412 653, width=8cm, clip]{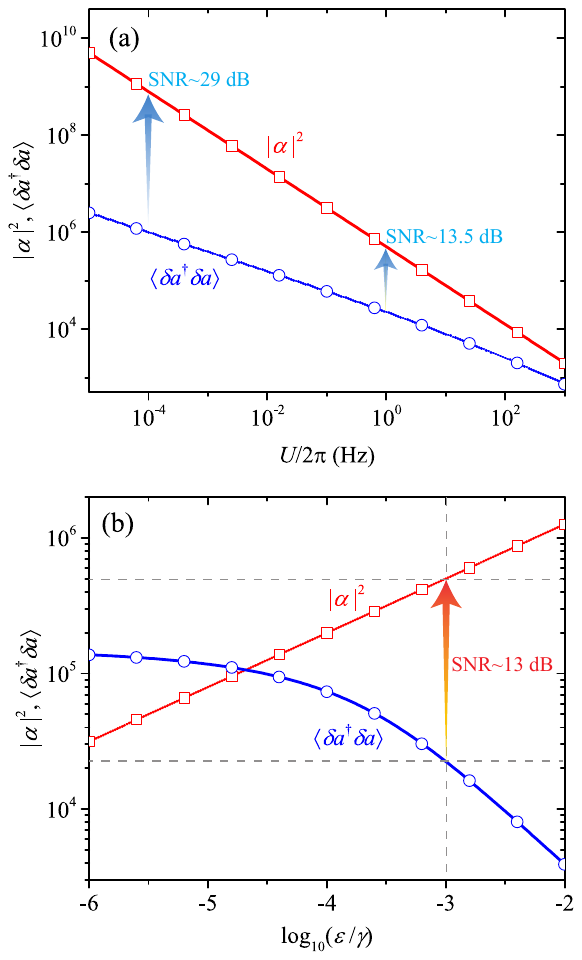}
		\caption{(a) The mean photon number $|\alpha|^2$ and the expectation of the correlated fluctuations $\langle\delta a^{\dagger}\delta a\rangle$ versus the Kerr coefficient $U/2\pi$, with the scaled amplitude of the single-photon drive \( \varepsilon / \gamma = 10^{-3} \). (b) $|\alpha|^2$ and $\langle\delta a^{\dagger}\delta a\rangle$ versus $\log_{10}(\varepsilon / \gamma)$ with \( U/2\pi = 1 \) Hz. The parameters used are $ \gamma/2\pi = 1$ GHz, $\Delta=0$, and \( G = \gamma/4 \).}   
		\label{fig4}
	\end{figure}
	
Technically, the mean photon number is given by $N=\langle a^{\dagger} a\rangle=|\alpha|^2 +\langle \delta a^{\dagger} \delta  a\rangle$, where $|\alpha|^2$ is the semiclassical signal and the correlation of the fluctuation operator $\langle \delta a^{\dagger} \delta  a\rangle$ can be regarded as the quantum noise. Thus $N=|\alpha|^2$ obtained based on the mean-field equations [Eqs.~(\ref{MF_eq1}) and (\ref{MF_eq2})] are credible under the condition that the semiclassical signal is much stronger than the quantum noise, i.e., the $|\alpha|^2 \gg \langle \delta a^{\dagger} \delta  a\rangle$. To characterize the effectiveness of the mean-field equations, we can introduce a signal-to-noise ratio (SNR) as
\begin{equation}
{\rm SNR}=10 \log_{10} \left(\frac{|\alpha|^2}{\langle \delta a^{\dagger} \delta  a\rangle}\right).
\end{equation}
As shown in the previous works~\cite{ZhangGQ2021PRB,XuYJ2024PRA} that the $\langle \delta a^{\dagger} \delta  a\rangle$ diverges when $G=G_0$, which is bad for sensing. How to enhance the SNR is an important question for sensing in practice.
Here, to avoid the divergence of $\langle \delta a^{\dagger} \delta  a\rangle$, we consider that the optical cavity mode is subjected to both single-photon and two-photon drives, and the correlation effect induced by the Kerr nonlinearity is also included in the following discussion. We will show that a high SNR can be obtained at the critical two-photon drive strength ($G=G_0$).

Based on the QLEs for the fluctuation operator $\delta a$ and $\delta a^{\dagger}$ [Eqs.~(\ref{QLE_f_eq1}) and (\ref{QLE_f_eq2})], we can obtain the dynamic equations for the expectation of the correlated fluctuations $\langle\delta a^{\dagger}\delta a\rangle$ and  $\langle\delta a \delta a\rangle$ as
\begin{equation}\label{noise1}
			\frac{d\left\langle \delta a^{\dagger} \delta a\right\rangle}{d t}=  -\gamma\left\langle \delta a^{\dagger} \delta a\right\rangle-2 i G'\left\langle \delta a^{\dagger} \delta a^{\dagger}\right\rangle+2 i G'^*\langle \delta a \delta a\rangle,
\end{equation}
	\begin{eqnarray}\label{noise2}
			\frac{d\langle \delta a \delta a\rangle}{d t}= &\left(-i 2\Delta''-\gamma\right)\langle \delta a \delta a\rangle-2 i G'\left(2\left\langle \delta a^{\dagger} \delta a\right\rangle+1\right)\nonumber \\
			& -i 2 U\left(2\left\langle \delta a^{\dagger} \delta a \delta a \delta a\right\rangle+\langle \delta a \delta a\rangle\right).
\end{eqnarray}	
These equations cannot be strictly solved due to the presence of the fourth-order correlation \( \langle \delta a^\dagger \delta a \delta a \delta a \rangle \). Under the weak Kerr nonlinearity condition ($U \ll \gamma$), the noise of the system can be treated as Gaussian noise. 	For the properties of the Gaussian noise~\cite{Ford1965JMP,Ford1988PRA}, the fourth-order correlation \( \langle \delta a^\dagger \delta a \delta a \delta a \rangle \) can be expressed as the sum of the products of second-order correlations as 
\begin{equation}\label{4thcorr}
\left\langle \delta a^{\dagger} \delta a \delta a \delta a\right\rangle= 3 \left\langle \delta a^{\dagger} \delta a \right\rangle \left\langle  \delta a \delta a\right\rangle.
\end{equation}
After substituting this into Eqs.~(\ref{noise1}) and (\ref{noise2}), the correlated fluctuations $\langle\delta a^{\dagger}\delta a\rangle$ and  $\langle\delta a \delta a\rangle$ can be obtained numerically.
The coefficients in Eqs.~(\ref{noise1}) and (\ref{noise2}) depend on both the Kerr coefficient $U$ and $\alpha$, and $\alpha$ can be tuned by the amplitude of single-photon drive $\varepsilon$. So the SNR can be justified by tuning the amplitude of the single-photon drive.
	
To investigate the effect of the system's Kerr coefficient \( U \) on $|\alpha|^2$ and $\langle \delta a^{\dagger} \delta  a\rangle$, we show them as functions of $U$ in Fig~\ref{fig4}(a). Fig.~\ref{fig4}(a) shows that, in the weak Kerr nonlinearity limit ($U\rightarrow 0$), both of $|\alpha|^2$ and $\langle \delta a^{\dagger} \delta  a\rangle$ are divergent, and the SNR rounds towards infinity. With the increase of the nonlinear coefficient, both the steady-state photon number and the quantum noise are suppressed, and most importantly the SNR also decreases gradually. It means when the Kerr coefficient is relatively large, the effect of the quantum noise becomes significant. One effective way to improve the SNR is enhancing the mean photon number and suppressing the quantum noise at the same time by enhancing single-photon drive.
	
We show the mean photon number $|\alpha|^2$ and the correlated fluctuations $\langle\delta a^{\dagger}\delta a\rangle$ as functions of the single-photon driving strength $\log_{10}(\varepsilon / \gamma)$ in Fig.~\ref{fig4}(b). When the driving strength is zero or very small (i.e., $\varepsilon<\gamma/10^5$), the quantum noise is much greater than the mean photon number. However, with the increase of the single-photon driving strength, the mean photon number of the system is enhanced, while the quantum noise is suppressed rapidly. 
When the driving strength is relatively high (i.e., $\varepsilon>\gamma/10^4$), the mean photon number of the system becomes significantly greater than the system noise. Specifically, the SNR is about 13 dB when the driving strength $\varepsilon=\gamma/10^3$, with the parameters used in this paper.
This indicates that the single-photon drive provides an effective way to suppress the quantum noise and improve the SNR. 
	
\section{Conclusion}\label{Concl}
	
In this paper, we have proposed a scheme for sensing the weak Kerr nonlinearity in an optical cavity, based on the CP for phase transition, by both single- and two-photon drives. 
We found that the mean photon number around the CP induced the two-photon drive sensitively depends on the Kerr coefficient in the optical cavity, and showed that the weak anharmonicity in the optical systems can be measured sensitively by detecting the mean photon number.
Moreover, we discussed the effect of the quantum noise and demonstrated that the single-photon drive provides an effective way to suppress the quantum noise and improve the SNR. 
This scheme can be applied to detecting different kinds of weak nonlinear interactions which can induce anharmonicity in optical cavities, such as the atom-resonator interactions, magnon-photon interactions, and the optomechanical interactions.

%\section{Acknowledgement}
\vskip 2pc \leftline{\bf Acknowledgement}

This work was supported by the Innovation Program for Quantum Science and Technology (Grant No.~2024ZD0301000), the National Natural Science Foundation of China (Grants No.~12064010, No.~12247105, and No.~12421005),
the science and technology innovation Program of Hunan Province (Grant No.~2022RC1203), 
and Hunan Provincial Major Sci-Tech Program (Grant No.~2023ZJ1010).

	\bibliography{ref}
	%\bibitem [{sup()}]{suppl}%
	%  \BibitemOpen
	%  \href@noop {} {}\bibinfo {note} {See Supplemental Material, which includes Refs.~[45-48], for
		%  details on the calculation of the ISF, the pseudo-Brownian scheme, the mean-square
		%  displacement and the orientational correlation function for different
		%  P\'eclet numbers, and the tube diameter scaling for highly entangled active
		%  solutions.}\BibitemShut {Stop}
		
\end{document}